\documentclass[a4paper,11pt]{amsart}
\usepackage{graphicx}
\begin{document}
\hyphenation{gra-vi-ta-tio-nal re-la-ti-vi-ty Gaus-sian
re-fe-ren-ce re-la-ti-ve gra-vi-ta-tion Schwarz-schild
ac-cor-dingly gra-vi-ta-tio-nal-ly re-la-ti-vi-stic pro-du-cing
de-ri-va-ti-ve ge-ne-ral ex-pli-citly des-cri-bed ma-the-ma-ti-cal
de-si-gnan-do-si coe-ren-za pro-blem gra-vi-ta-ting geo-de-sic
per-ga-mon cos-mo-lo-gi-cal gra-vity cor-res-pon-ding
de-fi-ni-tion phy-si-ka-li-schen ma-the-ma-ti-sches ge-ra-de
Sze-keres con-si-de-red}
\title[Hilbertian Repulsive Effect and Dark Energy]
{{\bf Hilbertian Repulsive Effect and Dark Energy}}

\author[Angelo Loinger]{Angelo Loinger}
\address{A.L. -- Dipartimento di Fisica, Universit\`a di Milano, Via
Celoria, 16 - 20133 Milano (Italy)}
\author[Tiziana Marsico]{Tiziana Marsico}
\address{T.M. -- Liceo Classico ``G. Berchet'', Via della Commenda, 26 - 20122 Milano (Italy)}
\email{angelo.loinger@mi.infn.it} \email{martiz64@libero.it}

\vskip0.50cm

\begin{abstract}
A \emph{repulsive} gravitational effect of general relativity
(\emph{without} cosmological term), which was pointed out by
Hilbert many years ago, could play a decisive role in the
explanation of the observational data concerning the
\emph{accelerated} expansion of the universe.
\end{abstract}

\maketitle



\normalsize

\vskip1.20cm \noindent \textbf{\emph{Introduction.}} -- In
previous papers \cite{1} we have illustrated some physical
consequences of a \emph{repulsive} gravitational effect, which was
evidenced by Hilbert many years ago \cite{2}. This effect comes
forth -- in particular regions -- in the instance (\emph{e.g.}) of
the Einsteinian gravitational field (\emph{without} cosmological
term) of a mass concentrated in a very small volume. In accord
with Einstein and Hilbert, we have considered the singularities
(both ``soft'' and ``hard'') of the metric tensor of the field of
a mass point as unphysical \emph{loci}, which however can give an
idea of the behaviour of potential $g_{jk}$, $(j,k=1,2,3,4)$,
generated by an \emph{extended} body. Now, the minimal radius of
an extended mass $M$ is equal to $(9/8)\,2m$, where $m\equiv
GM/c^{2}$, as it was demonstrated by Schwarzschild \cite{3} for a
homogeneous sphere of an incompressible fluid, and by Weinberg
\cite{4} for a sphere of a generic fluid. \emph{In our papers}
\cite{1} \emph{we have considered this case}.

\par In the present Note we show that the above Hilbertian effect
could play an important role in the problem of the \emph{dark
energy}. Our results can be regarded as a prolegomenon to a
realistic cosmological model that explains physically the nature
of the dark energy, by avoiding \emph{ad hoc} assumptions, as
(\emph{e.g.}) the recourse to a cosmological term, or to a
``quintessence'' \cite{5}.

\par
The behaviour of the geodesic lines of Schwarzschild manifold is
interesting. For a gravitating mass point, test-particles and
light-rays in \emph{radial} motions arrive at the space surface
$r=2m$ (we employ here the customary \emph{standard} coordinates)
with a velocity $v=\textrm{d}r / \textrm{d}t$ and an acceleration
$\textrm{d}v / \textrm{d}t$ \emph{equal to zero}. (Accordingly,
not even the tricking assumption that for the ``internal'' region
$r<2m$ the direction $\partial / \partial r$ acquires a temporal
meaning -- and $\partial / \partial t$ a radial meaning -- can
validate a \emph{physical} extension to $r<2m$ of the
``external'', $r>2m$, radial geodesics). On the contrary,
test-particles and light-rays arrive at the surface of a
gravitating sphere of radius $(9/8)\,2m$ with a velocity and an
acceleration that can be relatively small, but non-zero.

\par This Hilbertian effect is an \emph{objective} physical
property, which is independent of the coordinate system. The same
results are obtained, \emph{e.g.}, with the \emph{original}
Schwarzschild's coordinates (\cite{3} and \cite{6}), with
Brillouin's coordinates \cite{7}, with Fock's coordinates \cite{8}
-- and with any coordinate frame which does \emph{not} ``conceal''
the ``soft'' singularity. Thus, the well-known coordinates $u, v$
of Kruskal and Szekeres must be excluded, because the derivatives
$\partial u / \partial r$ and $\partial v / \partial r$, ($r$
standard), are singular at $r=2m$. An analogous exclusion holds
for the celebrated coordinate systems of Eddington-Finkelstein, of
Lema\^\i tre, of Synge \cite{9}, of Novikov.

\par The Hilbertian repulsion is particularly evident for the
\emph{circular} geodesics of Schwarzschild manifold. For the
material particles these geodesics are restricted by the following
inequalities (standard coordinates) \cite{2}:

\begin{equation} \label{eq:one}
r > \frac{3}{2}\cdot 2m \quad; \quad \frac{v}{c} <
\frac{1}{\sqrt{3}} \quad,
\end{equation}

where $v=c \,(m/r)^{1/2}$ is the (linear) velocity. For the
circular trajectories of the light-rays, the coordinate radius $r$
is equal to $(3/2) \,2m$, and the velocity $v$ is equal to $
c/\sqrt{3}$ \cite{2}. For small $r$'s the Einsteinian gravity acts
repulsively. Since $3/2$ $>$ $9/8$, we see that no circular
geodesic can touch the surface of a material sphere of physical
radius $(9/8)\, 2m$. (With standard coordinates, this radius
coincides with the coordinate radius, say $r_{c}$, of the sphere).

\par It is commonly emphasized that if we adopt, \emph{e.g.}, the
coordinates of Synge \cite{9} or of Kruskal-Szekeres, or of
Novikov, the geodesic lines penetrate the regions corresponding to
the region $r\leq 2m$: geodesic completeness. A physically
insignificant result, because -- as we have pointed out -- these
coordinate systems ``hide'' the ``soft'' singularity $r=2m$ -- and
are therefore unreliable. Moreover, in these global sets of
coordinates the gravitational field is \emph{non}-static. Synge,
Kruskal, Szekeres, Novikov have merely camouflaged a
stumbling-block.

\par Quite incomprehensibly, in the current literature Hilbert's
significant  and complete treatment of the Schwarzschild manifold
\cite{2} is ignored; of special importance is his investigation of
the relevant \emph{first integrals} of the equations of motion of
test-particles and light-rays, by means of which an intrinsic
characterization of the geodesic lines can be given. (An English
translation of Hilbertian memoirs on general relativity would be
highly desirable).

\vskip1.20cm \noindent \textbf{1.} -- Recent observations tell us
that: \emph{i}) the mass density (of visible and dark matter) of
the universe is \emph{circa} equal to $(1/3)$ of the critical
density of Friedmann model; \emph{ii}) the expansion of the
universe is monotonically increasing -- and in an
\emph{accelerated} way; \emph{iii}) the universe seems to be
spatially flat \cite{10}; \emph{iv}) most of the universe mass is
\emph{dark}; \emph{v}) the present value $H_{0}$ of Hubble's
function $H(t)$ is: $(65 \pm 10)$ km $\cdot$ s$^{-1}\,$Mpc$^{-1}$;
(1 Mpc $\stackrel{\wedge}{=} 3.26\times 10^{6}$ lyr).

\vskip1.20cm \noindent \textbf{2.} -- Many papers on \emph{dark
energy} contain improper considerations on Friedmann model, in
which the mass density $\varrho$ is \textbf{\emph{only}} composed
of the ``dust'' of the ``galaxy gas'' \cite{11}. If $\varrho_{c}:=
3H(t)^{2}/(8\pi G)$ is the
 critical density, we have $\Omega := \varrho / \varrho_{c} = 8\pi
 G\varrho /(3H^{2})$; the introduction of the cosmological term
 $\Lambda g_{jk}$ in Friedmann equations is certainly allowed, but
 it is not allowed to add to $\varrho$ a $\varrho_{v}:= c^{2}\Lambda / (8\pi
 G)$, interpreted -- from the standpoint of \emph{quantum} field
 theory -- as the mass-energy density of the \emph{vacuum}, and to
 consider an $\Omega_{v}:= 8\pi G \varrho_{v} / (3H^{2})$. Indeed,
 general relativity is a \emph{classical} theory, and therefore it
 does not admit to be hybridized with \emph{quantum} concepts. In
 particular, the concept of a vacuum energy is fully extraneous
 to GR.

 \par A cosmological term  can be introduced also in Newton
 theory, by substituting Poisson equation $\nabla^{2}\Phi =
 4\pi G\varrho$ with $\nabla^{2}\Phi - \Lambda \Phi=
 4\pi G\varrho$ \cite{12}. There exists a perfect
 \emph{isomorphism} between Friedmann model and the corresponding
 Newtonian model, both for $\Lambda =0$ and for $\Lambda\neq 0$
 \cite{11}.

 \par We remark also that the consideration of a mass density
 proportional to space curvature is conceptually unjustified.

 \par However, without the fictitious mass-energy density
 $\varrho_{v_{}}$  it seems impossible to explain -- with
 Friedamnn model -- the \emph{observational data}: indeed, since
 $\Omega\approx 1/3$, the assumption $\Omega_{_{v}}\approx 2/3$
 appears as arithmetically natural in order to have an
 $\Omega_{tot}= \Omega + \Omega_{v} = 1$, which would ensure the
 monotonic expansion of a spatially flat universe; the
 \emph{accelerated} motion would be the merit of the cosmological
 term.

 \par Experience counsels us to try the construction of a more
 adequate cosmological model.

 \vskip1.20cm \noindent \textbf{3.} -- A purely \emph{kinematic
 model} of universe is based on the well-known Hubble's relation:

\begin{equation} \label{eq:two}
v = cz = H_{0} \, r \quad;
\end{equation}

where $v=\textrm{d}r / \textrm{d}t$ is the recession velocity, $z$
is the red-shift $\triangle \lambda / \lambda_{0}$, $r$ is the
distance of the considered galaxy. Eq.(\ref{eq:two}) can be
written in a vectorial form:

\begin{equation} \label{eq:three}
 \textbf{v} = H_{o}\,\textbf{r} \quad;
\end{equation}

an observer in another galaxy at a distance $\textbf{r}'$ and a
velocity $\textbf{v}'$ relative to us
$(\textbf{v}'=H_{0}\,\textbf{r}')$ would find

\begin{equation} \label{eq:four}
\textbf{v} - \textbf{v}' = H_{0}\,(\textbf{r}-\textbf{r}') \quad,
\end{equation}

and this proves that the kinematic model describes a homogeneous
and isotropic universe. Under the condition $curl \,\textbf{v}=0$,
eq. (\ref{eq:three}) gives the only velocity field which assures
both homogeneity and isotropy.

\par Any reasonable cosmological model must have as a consequence
eqs. (\ref{eq:two}), (\ref{eq:three}), (\ref{eq:four}).

\vskip1.20cm \noindent \textbf{4.} -- Let us assume that the core
of a future cosmological model is a material sphere $\textbf{S}$
of mass $M$, whose radius is $(9/8)\,2m$, (with $m\equiv
GM/c^{^{2}}$), and consider its Schwarzschild manifold (\cite{3},
\cite{6}). The ``galaxy gas'' be composed of a spherically
symmetric swarm of test-particles, which at a given time abandon
abruptly the periphery of $\textbf{S}$ with a suitable
\emph{radial} velocity -- why? because ``\emph{Im Anfang war die
Tat''}! -- as in Friedmann model.

\par According to Hilbert's treatment (\cite{2}, \cite{1}), the
equation of the \emph{radial} geodesics can be written, in
standard coordinates:

\begin{equation} \label{eq:five}
\frac{1}{c^{2}} \, \frac{\textrm{d}^{2}r}{\textrm{d}t^{2}} -
\frac{3}{2} \, \frac{2m}{r(r-2m)}
\left(\frac{\textrm{d}r}{c\textrm{d}t} \right)^{2} +
\frac{m(r-2m)}{r^{3}} = 0^{}\quad;
\end{equation}

with the \emph{first integral}:

\begin{equation} \label{eq:six}
\left(\frac{\textrm{d}r}{c\textrm{d}t} \right)^{2} = \left(
\frac{r-2m}{r}\right)^{2}  + A \left( \frac{r-2m}{r}\right)^{3}
\quad,
\end{equation}

the constant $A$, which is equal to zero for the light-rays, is
negative for the material particles; it is: $\varepsilon \leq
|A|\leq 1$ , with $\varepsilon>0$ and \emph{ad libitum} small.
Eqs. (\ref{eq:five}) and (\ref{eq:six}) tell us that the
acceleration is negative (attractive gravity) or positive
(\emph{repulsive} gravity) where, respectively:

\begin{equation} \label{eq:seven}
\left| \frac{\textrm{d}r}{c\,\textrm{d}t} \right| <
\frac{1}{\sqrt{3}} \, \frac{r-2m}{r} \quad,
\end{equation}

\begin{equation} \label{eq:eight}
\left| \frac{\textrm{d}r}{c\,\textrm{d}t} \right| >
\frac{1}{\sqrt{3}} \, \frac{r-2m}{r}  \quad.
\end{equation}

Putting $x:=r/(2m)$ and $y:=(\textrm{d}r\ / c\textrm{d}t)^{2}$,
eq. (\ref{eq:six}) can be rewritten as follows:

\begin{equation} \label{eq:nine}
y(x) = \left( \frac{x-1}{x}\right)^{2}  \left( 1- |A| \,
\frac{x-1}{x}\right) \quad;
\end{equation}

the following nine figures give the diagrams of $y(x)$ for the
following values of $|A|$: $1;\, 0.9; \,0.8; \,0.7; \,2/3; \,0.5;
\,10^{-1}; \,10^{-3}; \,10^{-6}$. The last five diagrams represent
motions through \emph{everywhere}-repulsive regions. (Fig.10 gives
the $y(x)$ for light-rays, $A=0$).

\newpage
\begin{figure}[!hbp]
\begin{center}
\includegraphics[width=1.0\textwidth]{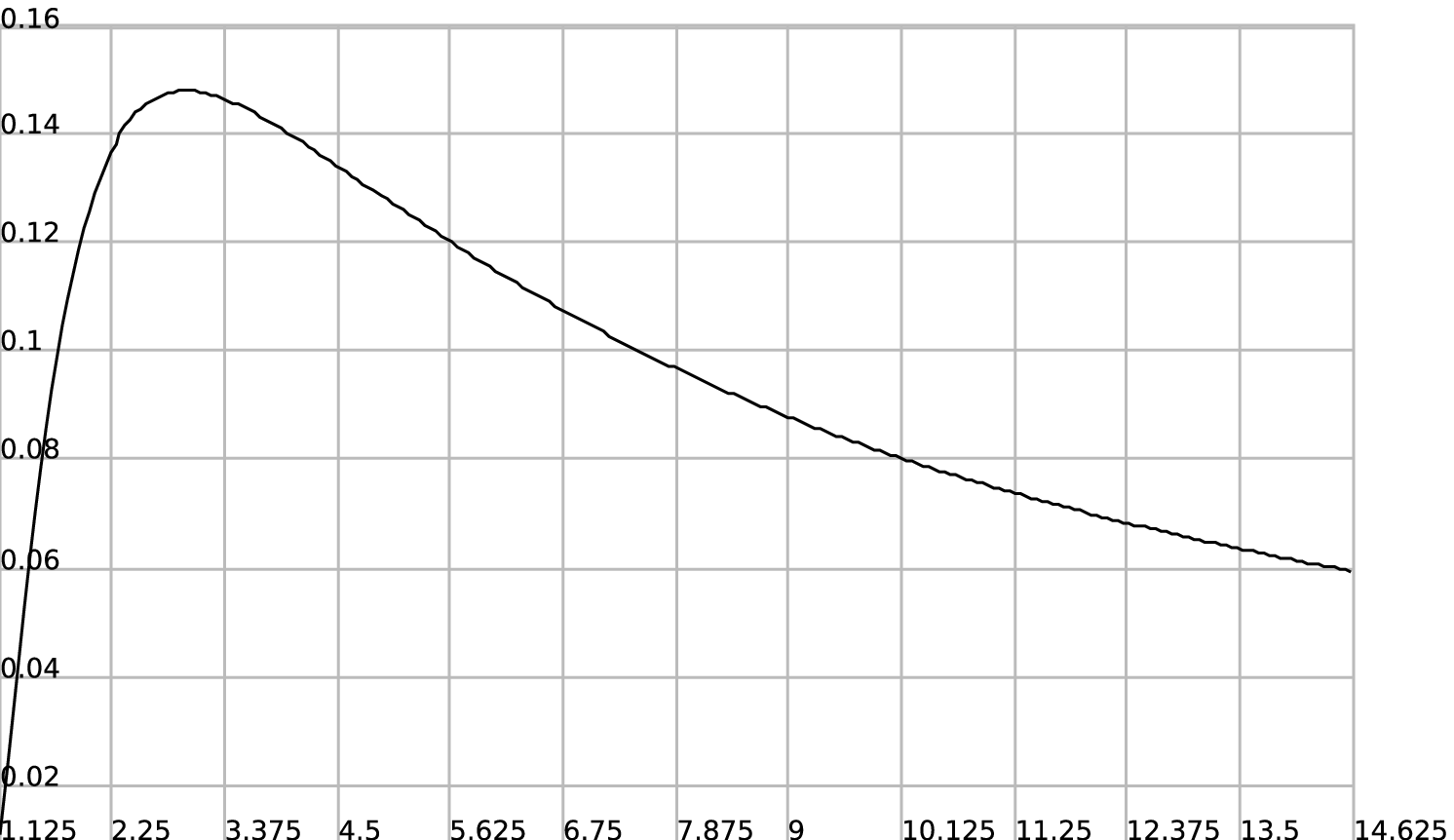}
\caption{Diagram of $y(x)=[(x-1)/x]^{2}[1-(x-1)/x]$ for some
values of $x$; $(9/8)\leq x <+\infty$; $\max(3.0;4/27)$;
$[y(9/8)]^{1/2}=2\sqrt2 /27$.} 
\vskip1.00cm
\includegraphics[width=1.0\textwidth]{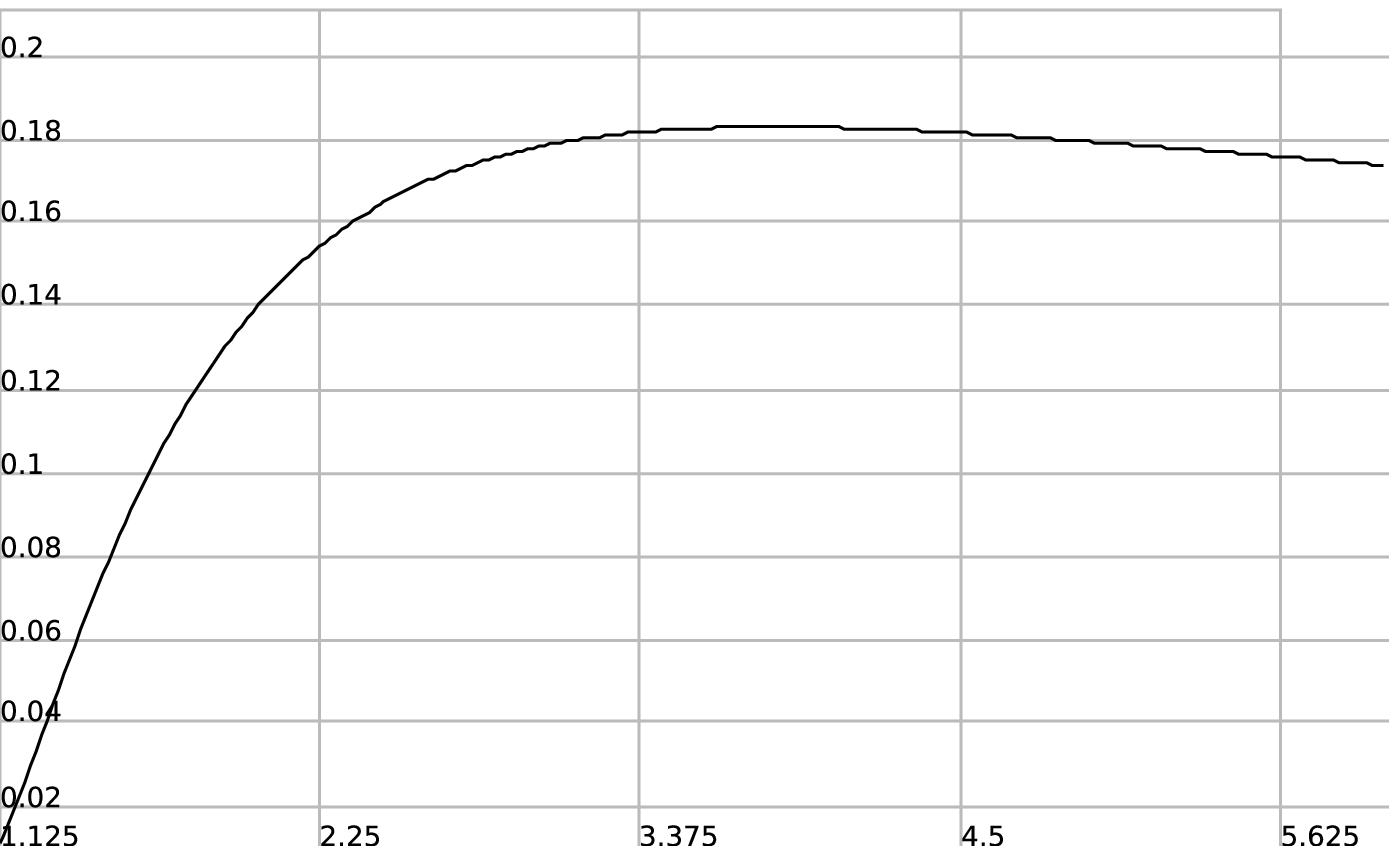}
\caption{Diagram of $y(x)=[(x-1)/x]^{2}[1-0.9*(x-1)/x]$ for some
values of $x$; $(9/8)\leq x <+\infty$; $\max(3.75;0.182844)$;
$[y(9/8)]^{1/2}=0.105409$.}
\end{center}
\end{figure}

\newpage
\begin{figure}[!hbp]
\begin{center}
\includegraphics[width=1.0\textwidth]{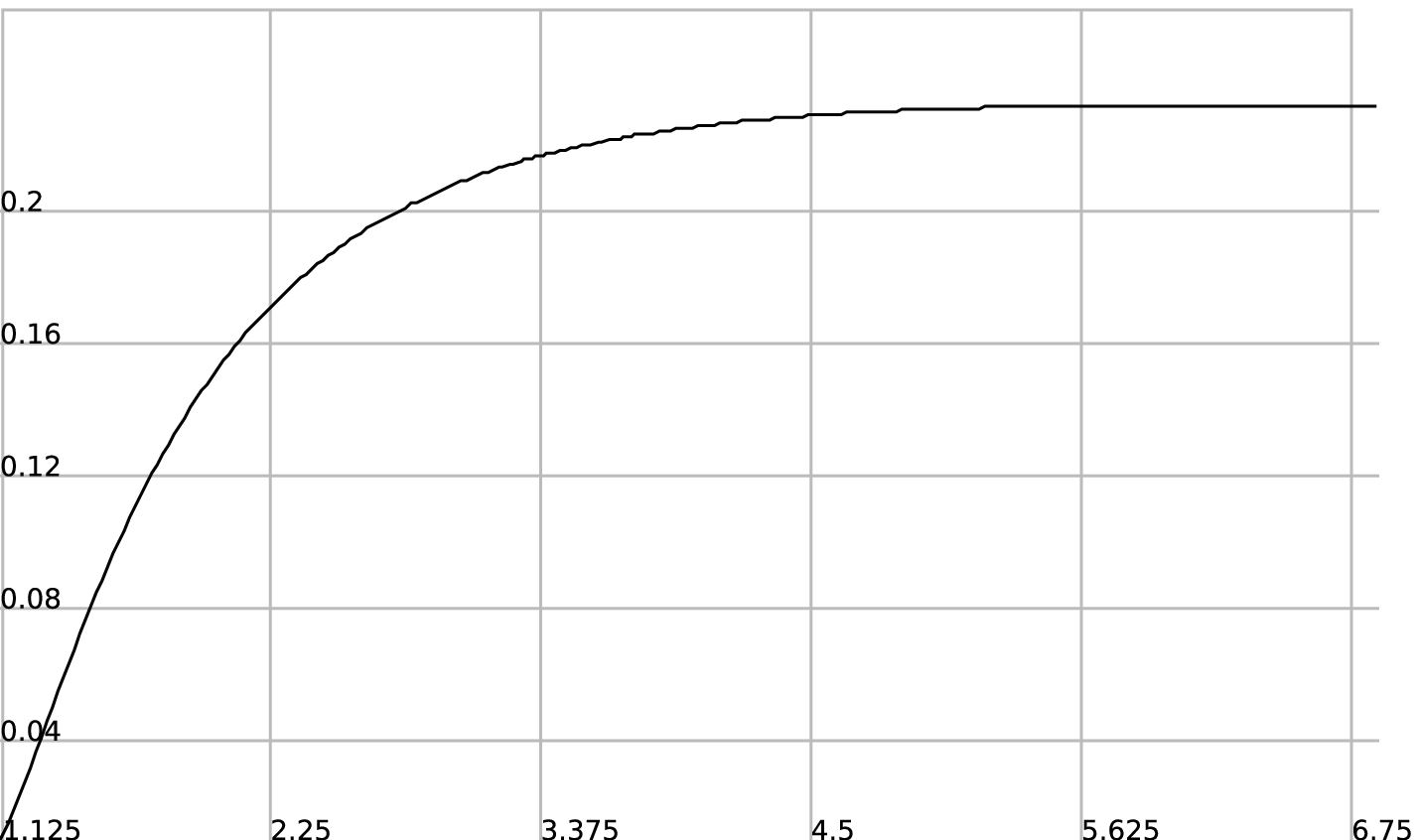}
\caption{Diagram of $y(x)=[(x-1)/x]^{2}[1-0.8*(x-1)/x]$ for some
values of $x$; $(9/8)\leq x <+\infty$; $\max(6.0;0.231481)$;
$[y(9/8)]^{1/2}=0.106058$.}
\vskip1.00cm
\includegraphics[width=1.0\textwidth]{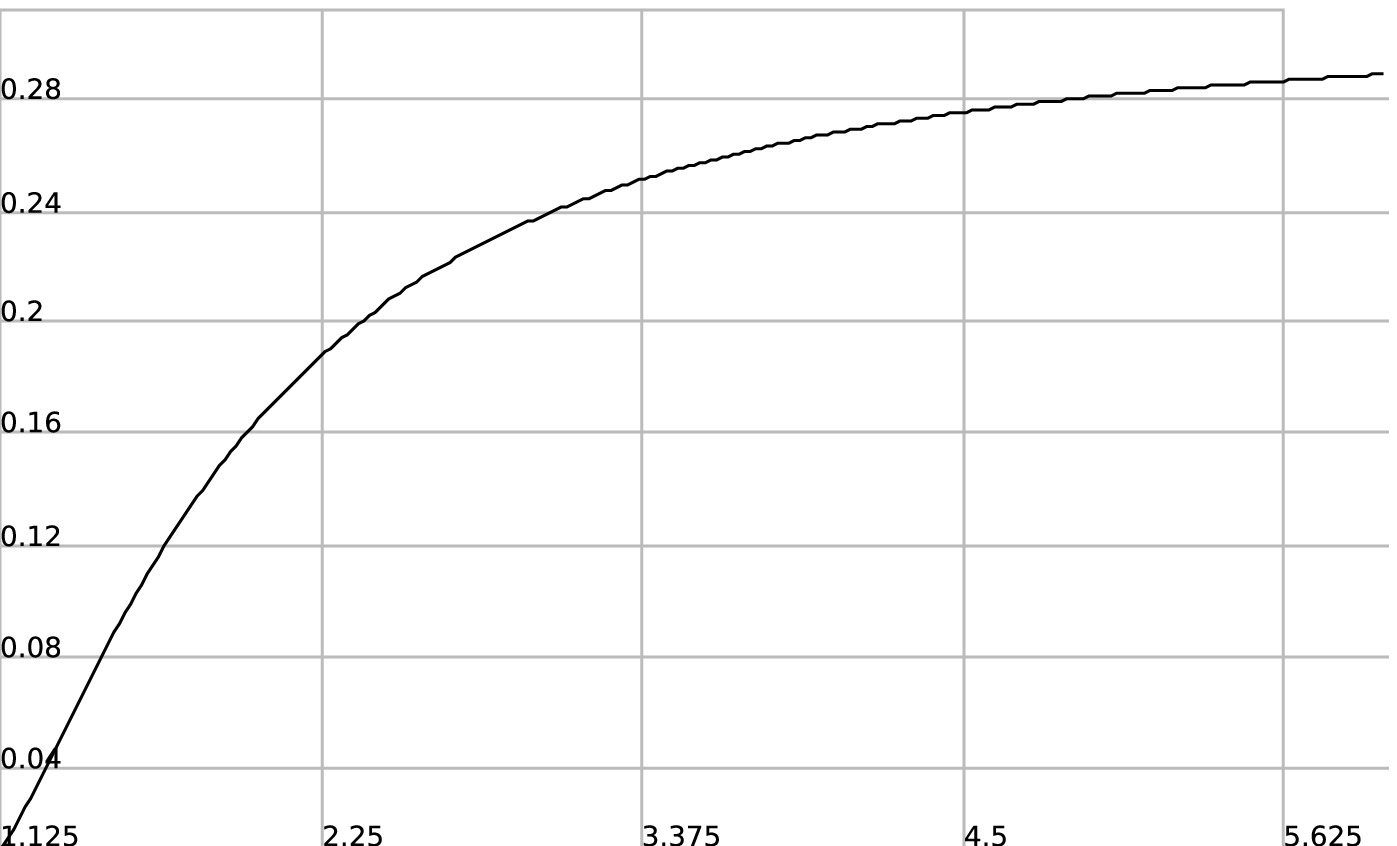}
\caption{Diagram of $y(x)=[(x-1)/x]^{2}[1-0.7*(x-1)/x]$ for some
values of $x$; $(9/8)\leq x <+\infty$; $\max(21.0;0.302343)$;
$[y(9/8)]^{1/2}=0.106703$.}
\end{center}
\end{figure}

\newpage
\begin{figure}[!hbp]
\begin{center}
\includegraphics[width=1.0\textwidth]{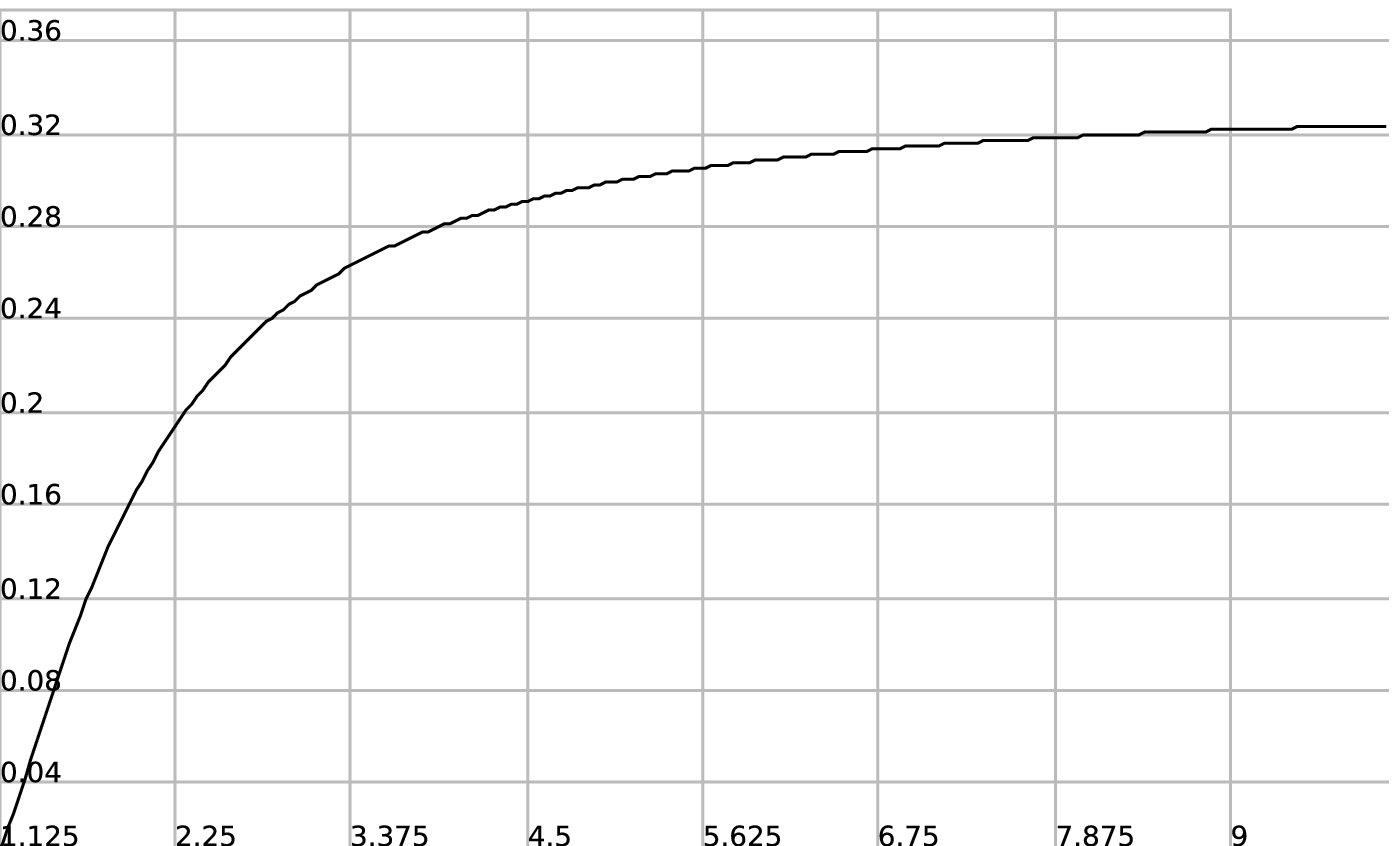}
\caption{Diagram of $y(x)=[(x-1)/x]^{2}[1-(2/3)*(x-1)/x]$ for some
values of $x$; $(9/8)\leq x <+\infty$; $\max(+\infty,1/3)$;
$[y(9/8)]^{1/2}=0.106917$.}
\vskip1.00cm
\includegraphics[width=1.0\textwidth]{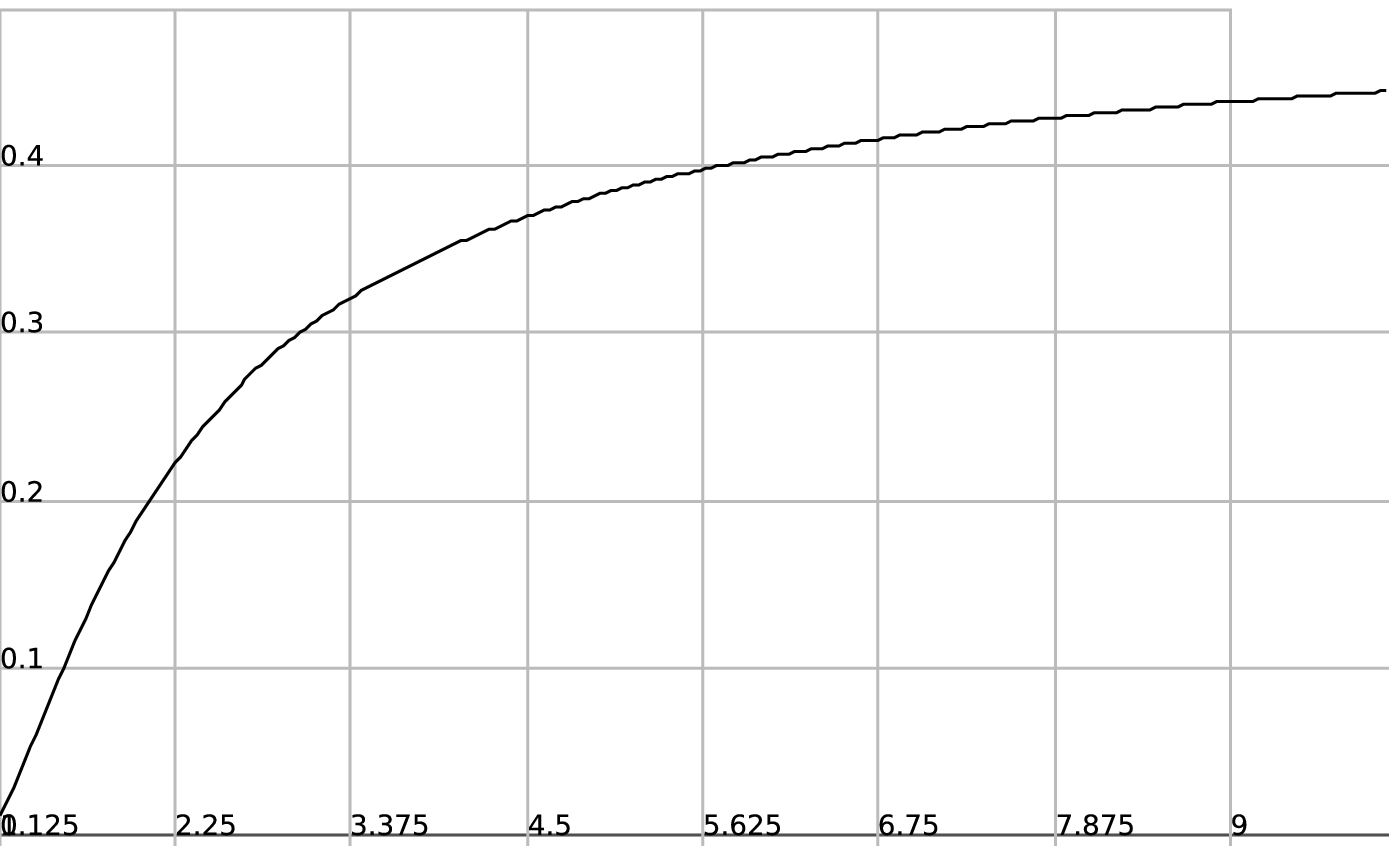}
\caption{\newline Diagram of $y(x)=[(x-1)/x]^{2}[1-0.5*(x-1)/x]$
for some values of $x$; $(9/8)\leq x <+\infty$;
$\max(+\infty;0.5)$; $[y(9/8)]^{1/2}=0.107981$.}
\end{center}
\end{figure}

\newpage
\begin{figure}[!hbp]
\begin{center}
\includegraphics[width=1.0\textwidth]{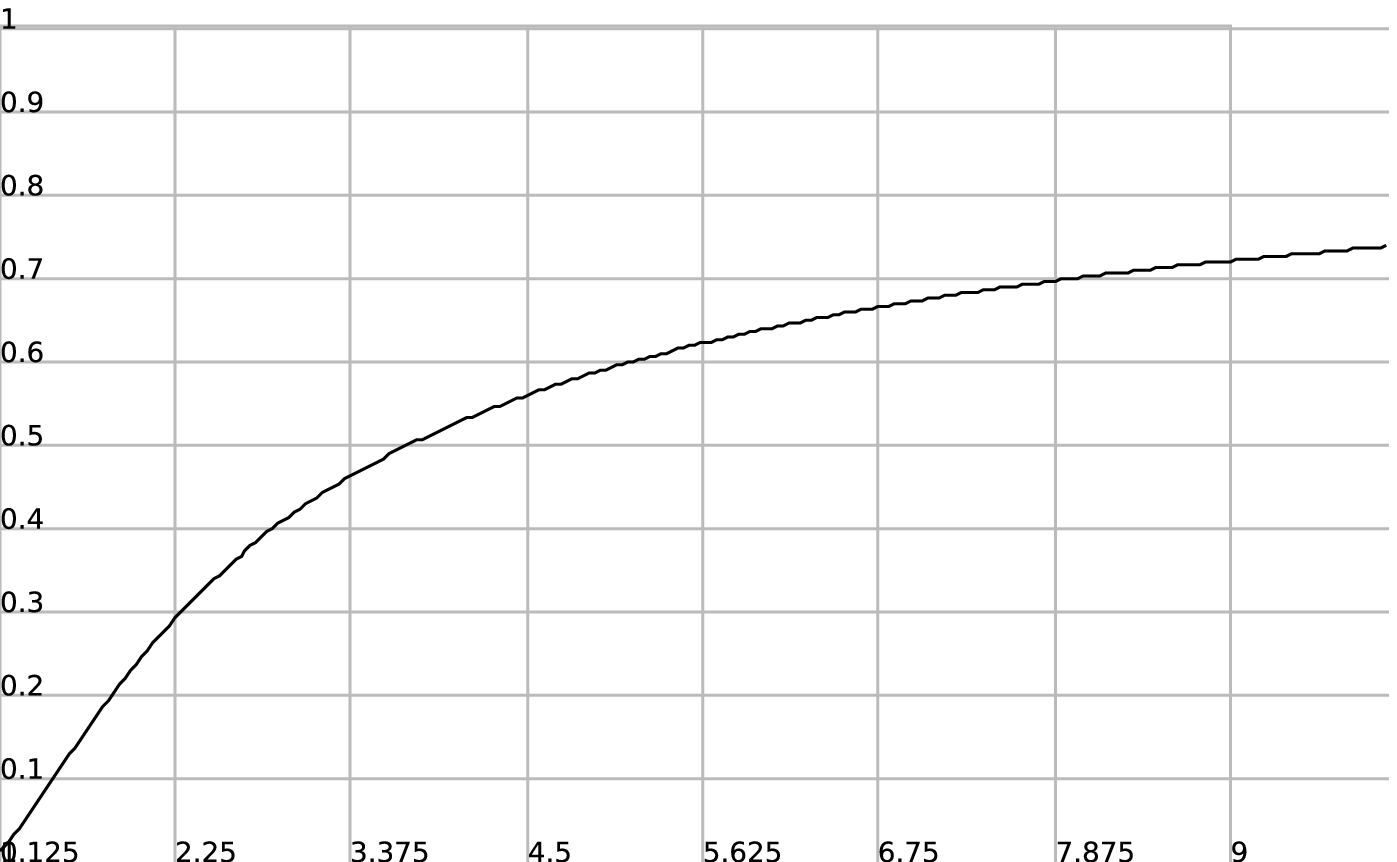}
\caption{\newline Diagram of
$y(x)=[(x-1)/x]^{2}[1-10^{-1}*(x-1)/x]$ for some values of $x$;
$(9/8)\leq x <+\infty$; $\max(+\infty;0.9)$;
$[y(9/8)]^{1/2}=0.110492$.}
\vskip1.00cm
\includegraphics[width=1.0\textwidth]{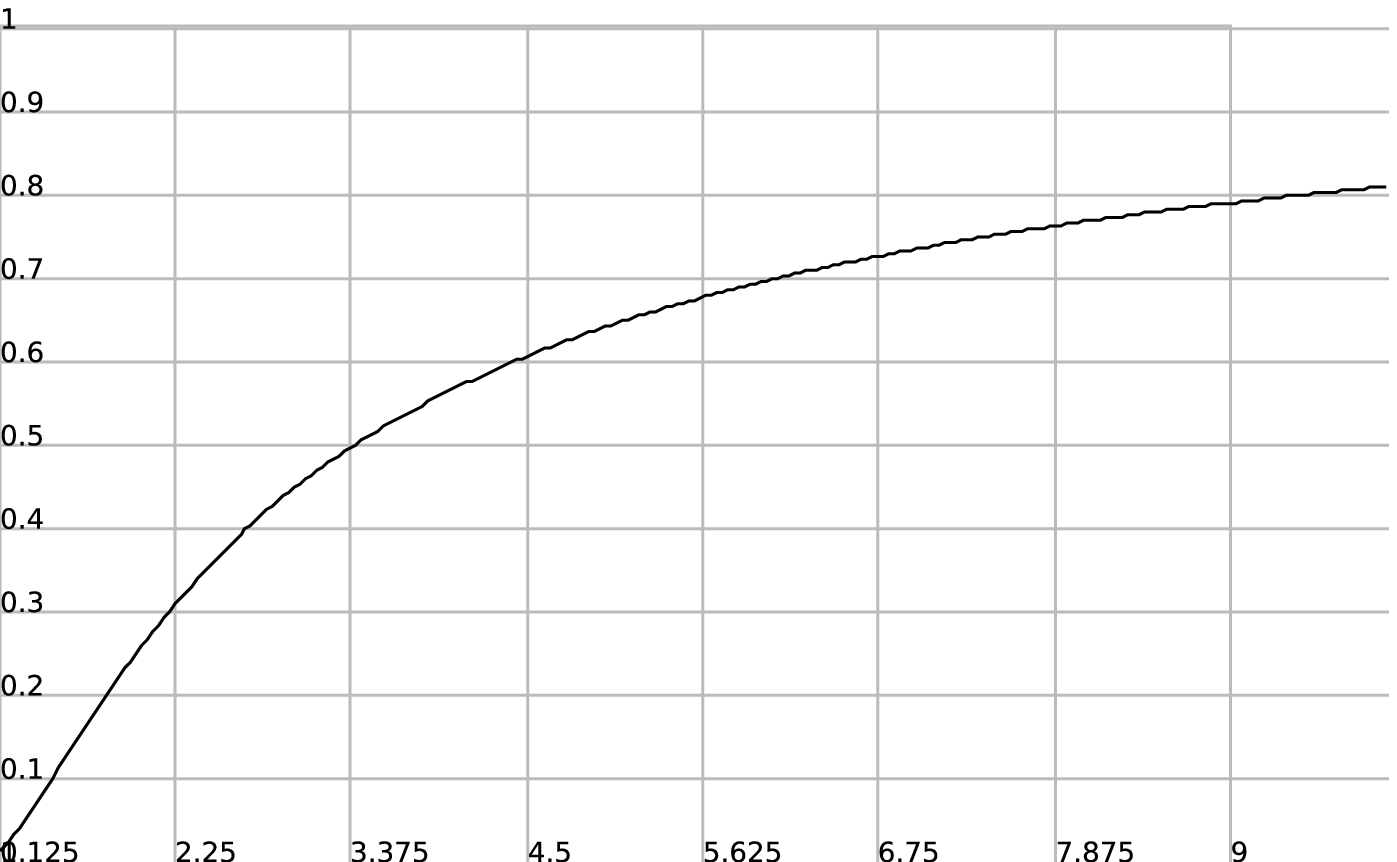}
\caption{\newline Diagram of
$y(x)=[(x-1)/x]^{2}[1-10^{-3}*(x-1)/x]$ for some values of $x$;
$(9/8)\leq x <+\infty$; $\max(+\infty;1-10^{-3})$;
$[y(9/8)]^{1/2}=0.111105$.}
\end{center}
\end{figure}

\newpage
\begin{figure}[!hbp]
\begin{center}
 \vskip1.00cm
\includegraphics[width=1.0\textwidth]{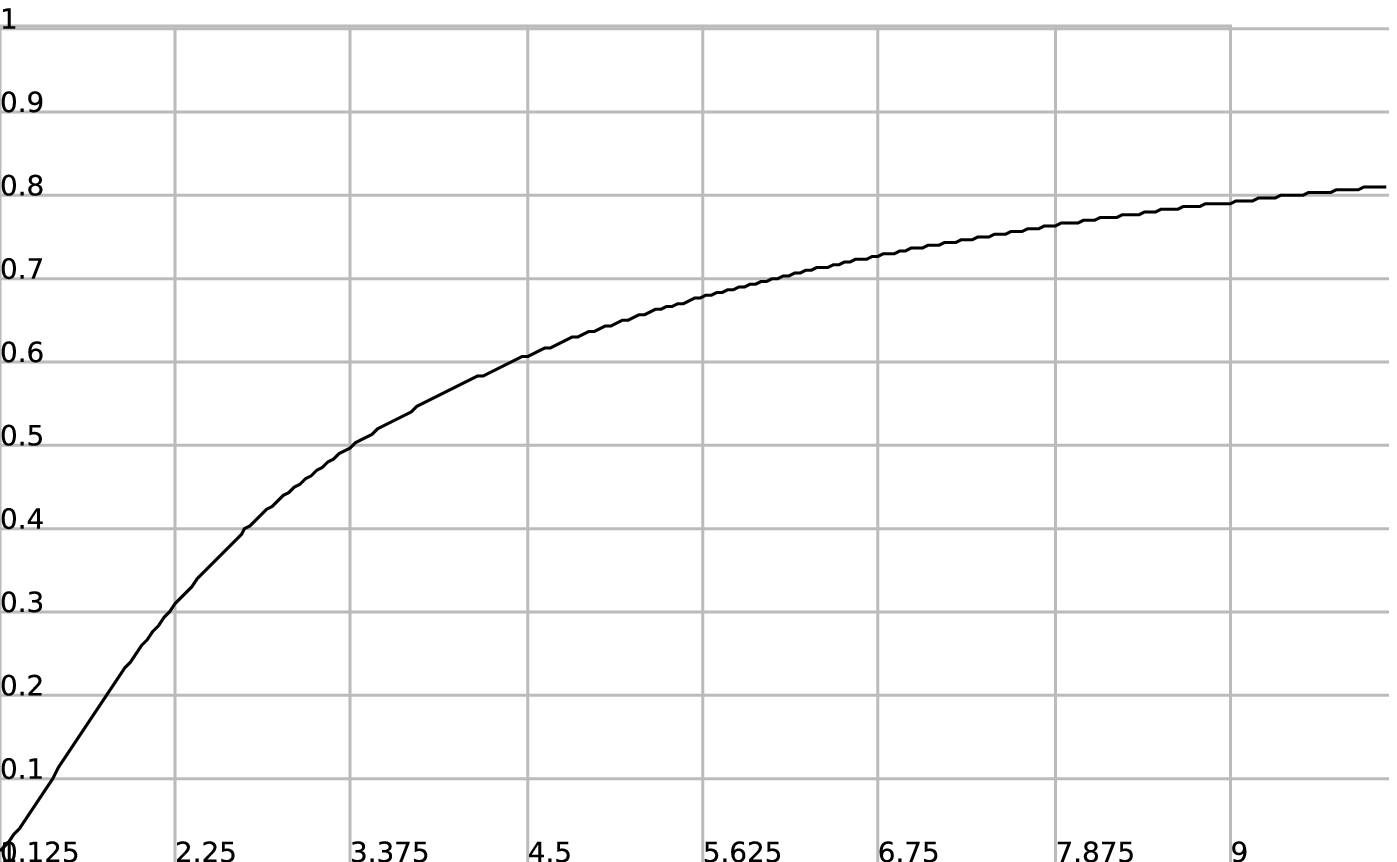}
\caption{\newline Diagram of
$y(x)=[(x-1)/x]^{2}[1-10^{-6}*(x-1)/x]$ for some values of $x$;
$(9/8)\leq x <+\infty$; $\max(+\infty;1-10^{-6})$;
$[y(9/8)]^{1/2}=0.111111$.}
\vskip1.00cm
\includegraphics[width=1.0\textwidth]{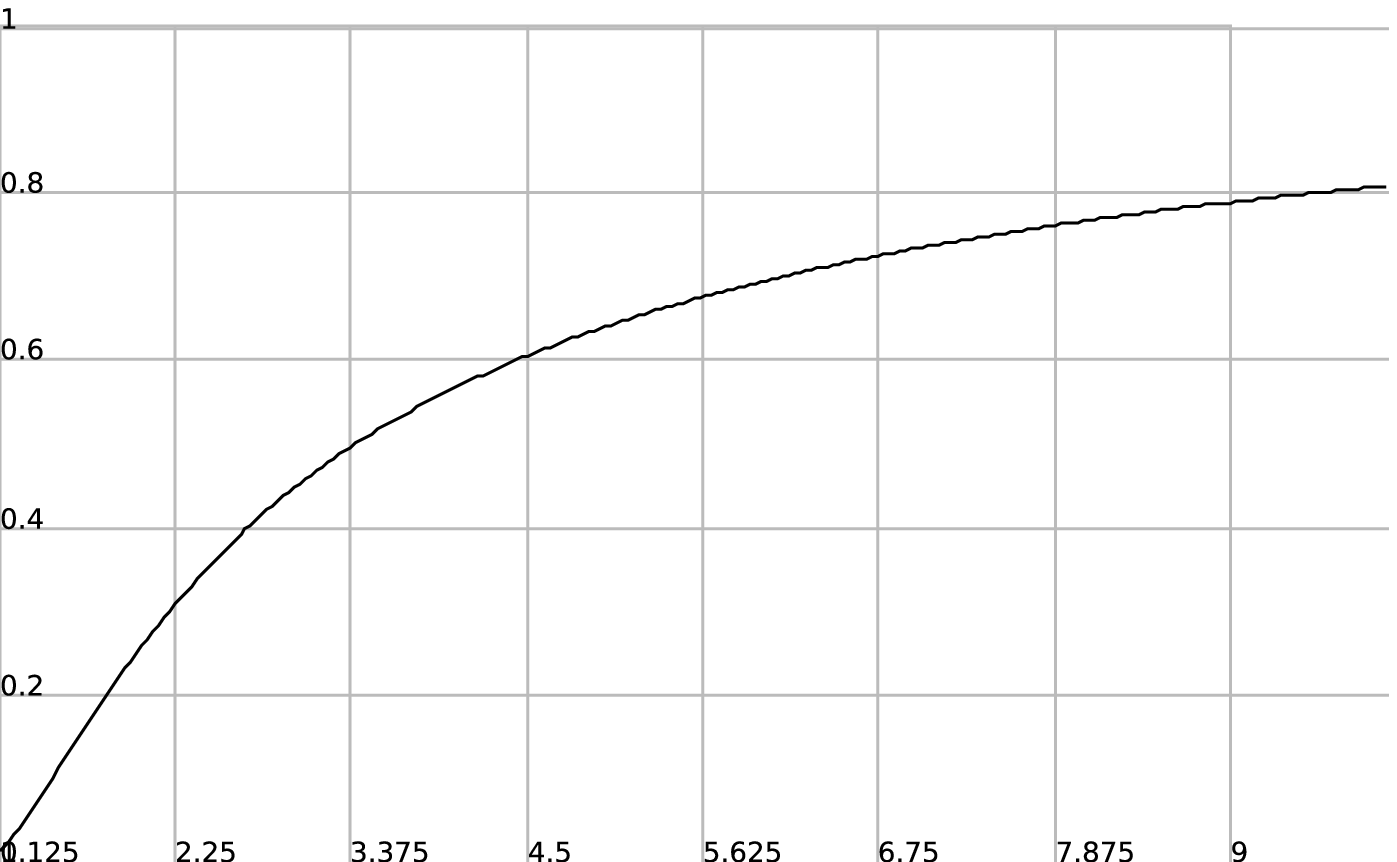}
\caption[5bis]{Diagram of $y(x)=[(x-1)/x]^{2}$ for some values of
$x$; $(9/8)\leq x <+\infty$; $\max(+\infty;1.0)$;
$[y(9/8)]^{1/2}=1 /9$.}
\end{center}
\end{figure}

\newpage
\noindent \textbf{5.} -- In Friedmann model with $\Lambda=0$,
there are relations among the scale factor $F(t)$, its derivatives
$\dot{F}(t)$ and $\ddot{F}(t)$, the mass density $\varrho(t)$, the
constant $\zeta$ $(=-1, \,0, \,+1)$ of space curvature. Experience
determines $\varrho(t_{0})$ and $H_{0}=H(t_{0})$, where $t_{0}$ is
the present time, and $H(t)= \dot{F}(t)/F(t)$ is Hubble's
function.

\par In our skeleton of cosmological model there are relations
among the radial coordinate $r(t)$, its derivatives $\dot{r}(t)$
and $\ddot{r}(t)$, the mass $M$ of the gravitating centre of
radius $(9/8)\,2m$, the integration constant $A$; for us the
function $\varrho(t)$ is theoretically free. Of course,
$H(t)=\dot{r}(t)/r(t)$, $A$ and $m$ are free parameters.
Experience determines $H(t_{0})=\dot{r}(t_{0})/r(t_{0})$ and
$\varrho(t_{0})$.

\par If $d(t)$ is the distance between two test-particles (two
galaxies), we have

\begin{equation} \label{eq:ten}
\textrm{d}(t)= \textrm{const}\cdot r(t) \quad,
\end{equation}

for which
\begin{equation} \label{eq:eleven}
\dot{\textrm{d}}(t) = \textrm{d}(t) \, \frac{\dot{r}(t)}{r(t)} =
\textrm{d}(t) \, H(t) \quad,
\end{equation}

and we see that the fundamental relation of the kinematic model is
satisfied.

 \vskip1.20cm \noindent \textbf{6.} -- The galaxy
gas and the gravitating body, hard core of the model, can be
composed, partially or even entirely, of dark matter.

\vskip1.20cm \noindent \textbf{7.} -- The main result of paper
\cite{10} is the following: the temperature anisotropies $\Delta
T/T$ of CMB (cosmic microwave background) tell us that the
universe is spatially flat ($\zeta =0$ in Friedmann model).
However, this assertion is not purely empirical, since depends
heavily on the assumed adequacy of the inflationary model. In our
scheme the universe has the spacetime curvature of Schwarzschild
manifold, which decreases gradually going away from the
gravitating centre.

\vskip1.20cm \noindent \textbf{\emph{Conclusion}}. The present
paper intends simply to draw the attention on the possible role of
Hilbertian repulsive effect (\cite{2}, \cite{1}) for the
explanation of the \emph{accelerated} expansion of the universe.

\par A final remark. Our insistence on the field generated by an
\emph{extended} gravitating centre, with a minimal radius
$(9/8)\,2m$, has the aim to show how useless are all the attempts
to give a physical meaning to singular geometrical \emph{loci}.

\newpage
\vskip2.00cm
\begin{center}
\noindent \small \emph{\textbf{APPENDIX}}
\end{center}
\normalsize \noindent \vskip0.80cm

\par Hilbert's eqs. (41), (42), (43) -- see \cite{2} -- give
immediately ($p$ is an affine evolution-parameter):

\begin{equation} \label{eq:twelve}
\left[\frac{\textrm{d}r(p)}{\textrm{d}p} \right]^{2} = C^{2} -
\left(1-\frac{2m}{r(p)}\right) \left(|A|+ \frac{B^{2}}{r^{2}(p)}
\right) \quad,
\end{equation}

where $A=-|A|$, $B$, $C$ are integration constants; $A$ and $C$
are numbers, $[B]=[\textrm{LENGTH}]$. Since
$r^{2}\textrm{d}\varphi (p) / \textrm{d}p = B$, we see that $B$
gives the value of the angular momentum in the plane of motion
$\vartheta=\pi /2$. The meaning of constant $C$ is given by
Hilbert's eq. (43), (H. puts $c=1$):

\begin{equation} \label{eq:thirteen}
\frac{r(p)-2m}{r(p)} \,\, \frac{c\, \textrm{d}t(p)}{\textrm{d}p} =
C
 \quad;
\end{equation}

with a trivial change of the parameter $p$, the constant $C$ can
be normalized to 1.

\par We see that Hilbert utilizes five \emph{first integrals}:
those characterized by $A$, $B$, $C$, plus the two remaining
components of angular momentum, that are implicitly determined by
the choice $\vartheta=\pi /2$. The \emph{physical} first integrals
are obviously four: the energy and the components of the angular
momentum. The energy constant is $A$. The constant $C$ owes its
existence to the introduction of the auxiliary parameter $p$.

\par Hilbert employed an affine $p$, in lieu of the proper time
$\tau$ -- or of $s=c\tau$ --, in order that his equations hold
\emph{both} for material particles \emph{and} light-rays. Most
authors employ $s=c\tau$ as an evolution parameter for the motions
of the \emph{test-particles}. Then, the constant $A$ has the
unique value $-1$, the constant $B$ maintains its meaning, and the
meaning of $C$ is given by the following equation:

\begin{equation} \label{eq:fourteen}
\frac{r(s)-2m}{r(s)} \,\,  \frac{c\,\textrm{d}t(s)}{\textrm{d}s} =
C
 \quad,
\end{equation}

from which we see that $C$ gives the value of
$g_{44}(s)=-(r(s)-2m)/r(s)$ for $s=s_{0}$, when the particle
begins its motion and $[\textrm{d}ct(s)/\textrm{d}s]_{s=s_{0}}=1$.

\par When the evolution parameter is $p$, the constant $A$ is
dynamically important, and $C$ plays only a limited role.
\emph{Vice-versa}, when the evolution parameter is $s$, the
constant $C$ is dynamically important, while $A$ plays only a
limited role.

\par The description of the motions by means of \emph{coordinate time
t} is the most adequate from the standpoint of the physical
meaning.


\newpage

\vskip1.80cm \small
\begin{thebibliography}{99}

\bibitem{1}
A. Loinger and T. Marsico: \emph{a}) \emph{arXiv:0706.3891 v3}
$[$physics.gen-ph$]$ 16 Jul 2007; \emph{b}) \emph{ibid.}:
\emph{0710.3927 v1} $[$physics.gen-ph$]$ 21 Oct 2007; \emph{c})
\emph{ibid.}: \emph{0711.4997 v3} $[$physics.gen-ph$]$ 22 Dec
2007.

\bibitem{2}
D. Hilbert, \emph{Mathem. Annalen}, \textbf{92} (1924) 1; also in
\emph{Gesammelte Abhandlungen}, Dritter Band (Springer-Verlag,
Berlin) 1935, p.258.

\bibitem{3}
K. Schwarzschild, \emph{Berl. Ber.}, (1916) 424; an English
version in \emph{arXiv:physics/9912033} (December 16th, 1999).

\bibitem{4}
S. Weinberg, \emph{Gravitation and Cosmology etc.} (Wiley and
Sons, New York, \emph{etc}.) 1972, Chapt.\textbf{11}, sect.
\textbf{6}.

\bibitem{5} See, \emph{e.g.}: Y. Wang and P. Mukherjee,
\emph{arXiv:astro-ph/0703780} -- 21 March 2007; A. Balbi and C.
Quercellini, \emph{arXiv:0704.2350 v3} $[$astro-ph$]$ 12 Nov 2007.
And references therein.

\bibitem{6}
K. Schwarzschild, \emph{Berl.Ber.}, (1916) 189; an English
translation in: \emph{i}) \emph{arXiv:physics/9905030} (May 12th,
1999); \emph{ii}) \emph{Gen. Rel. Grav.}, \textbf{35} (2003) 951.

\bibitem{7}
M. Brillouin, \emph{Journ. Phys. Rad.}, \textbf{23} (1923) 43; an
English version in \emph{arXiv:physics/0002009} (February 3rd,
2000).

\bibitem{8}
V. Fock, \emph{The Theory of Space, Time and Gravitation},
(Pergamon Press, Oxford, \emph{etc}.) 1964, sect. \textbf{57}.

\bibitem{9} J.L. Synge, \emph{Proc. Roy. Ir. Acad.}, \textbf{53A}
(1950) 83.

\bibitem{10} W. Hu and S. Dodelson, \emph{arXiv:astro-ph/0110414 v1} -- 18 Oct
2001, and in \emph{Annu. Rev. Astron. and Astrophys.} 2002. And
references therein.

\bibitem{11}
A. Uns\"old and B. Baschek, \emph{The New Cosmos}, Fourth
Completely Revised Edition (Springer-Verlag, Berlin, \emph{etc.})
1991, sect. \textbf{5.9}; A. Loinger, \emph{arXiv:physics/0504018
v1} -- 3 April 2005. And references therein.

\bibitem{12}
W. Pauli, \emph{Teoria della Relativit\`a} (Boringhieri, Torino)
1958, p.271; and the reference therein.

\end{thebibliography}
\end{document}